\documentclass[useAMS,usenatbib, dvipdfmx]{mn2e}
\usepackage{graphicx}
\usepackage{amsmath, amssymb, amsfonts}
\usepackage{color}
\usepackage{url, hyperref}

\newcommand{\be}{\begin{eqnarray}}
\newcommand{\ee}{\end{eqnarray}}

\title[]{On the Systematic Errors of Cosmological-Scale Gravity Tests
using Redshift Space Distortion: Non-linear Effects and the Halo Bias}
\author[T. Ishikawa et al.] 
{Takashi Ishikawa$^{1}$\thanks{E-mail:ishikawa@kusastro.kyoto-u.ac.jp},  
Tomonori Totani$^{1,2}$, 
Takahiro Nishimichi$^{3,4}$, 
Ryuichi Takahashi$^{5}$, 
\newauthor 
Naoki Yoshida$^{3,6}$ 
and Motonari Tonegawa$^{1,2}$
\\
$^{1}$Department of Astronomy, Kyoto University, Kitashirakawa-Oiwake-cho, Sakyo-ku, Kyoto 606-8502, Japan\\
$^{2}$Department of Astronomy, The University of Tokyo, 7-3-1 Hongo, Bunkyo-ku, Tokyo 113-0033, Japan\\
$^{3}$Kavli Institute for the Physics and Mathematics of the Universe, 
Todai Institutes for Advanced Study, The University of Tokyo,\\
\, Kashiwa, Chiba 277-8583 Japan (Kavli IPMU, WPI)\\
$^{4}$Institut d???Astrophysique de Paris, 98 bis boulevard Arago, 75014 Paris, France\\
$^{5}$Faculty of Science and Technology, Hirosaki University, 3 bunkyo-cho, Hirosaki, Aomori, 036-8561, Japan\\
$^{6}$Department of Physics, The University of Tokyo, 7-3-1 Hongo, Bunkyo-ku, Tokyo 113-0033, Japan\\
}

\begin{document}

\date{Accepted --. Received --; in original form --}

\pagerange{\pageref{firstpage}--\pageref{lastpage}} \pubyear{--}

\maketitle

\label{firstpage}

\begin{abstract}%
  Redshift space distortion (RSD) observed in galaxy redshift surveys
  is a powerful tool to test gravity theories on cosmological scales,
  but the systematic uncertainties must carefully be examined for
  future surveys with large statistics.  Here we employ various
  analytic models of RSD and estimate the systematic errors on
  measurements of the structure growth-rate parameter, $f\sigma_8$,
  induced by non-linear effects and the halo bias with respect to the
  dark matter distribution, by using halo catalogues from 40
  realisations of $3.4 \times 10^8$ comoving $h^{-3}$Mpc$^3$
  cosmological N-body simulations. We consider hypothetical redshift
  surveys at redshifts $z\!=\!0.5$, 1.35 and 2, and different minimum
  halo mass thresholds in the range of $5.0 \times 10^{11}$--$2.0
  \times 10^{13}\,h^{-1} \mathrm{M_\odot}$.  We find that the
  systematic error of $f\sigma_8$ is greatly reduced to $\sim\!$ 5 per
  cent level, when a recently proposed analytical formula of RSD that
  takes into account the higher-order coupling between the density and
  velocity fields is adopted, with a scale-dependent parametric bias
  model. Dependence of the systematic error on the halo mass, the
  redshift, and the maximum wavenumber used in the analysis is
  discussed.  We also find that the Wilson-Hilferty transformation is
  useful to improve the accuracy of likelihood analysis when only a
  small number of modes are available in power spectrum measurements.
\end{abstract}


\section{Introduction}

Many observational facts suggest that our universe is now in the 
period of accelerated expansion but its physical origin is yet to be 
understood \citep{Riess1998, Perlmutter1999, Spergel2003, Tegmark2004}. 
This might be a result of an exotic form of energy with negative 
pressure that should be added to the right-hand-side of the Einstein 
equation as the cosmological constant $\Lambda$, or more generally a 
time varying dark energy term. Another possibility is that gravity 
is not described by the Einstein equation on cosmological 
scales. Therefore observational tests of gravity theories on 
cosmological scales are important, and the redshift space distortion 
(RSD) effect observed in galaxy redshift surveys gives such a 
test. RSD is distortion of a galaxy distribution in redshift space 
caused by peculiar motions of the galaxies (see \citealt{Hamilton1998} 
for a review). The magnitude of this effect is expressed by the 
anisotropy parameter $\beta\!=\! f/b$ at the linear level 
\citep{Kaiser1987}, where $f\!=\!\mathrm{d} \ln \delta/ \mathrm{d} \ln a$ 
is the linear growth rate of the fractional density fluctuations 
$\delta$, $a$ the scale factor of the universe, and $b$ the galaxy 
bias with respect to the matter distribution. This is simply a result 
of the mass continuity that relates the growth rate and the velocity 
of large-scale systematic infall motion, and thus is always valid 
regardless of gravity theories.  When the galaxy bias is independently 
measured, one can derive the parameter $f$.  When the galaxy bias is 
unknown, we can still measure the combination of $f\sigma_8$ using 
the observed fluctuation amplitude of the galaxy density field, where 
$\sigma_8$ is the rms amplitude of the mass fluctuations on comoving 
$8\,h^{-1}\mathrm{Mpc}$ scale. 

A number of measurements of the growth rate have been reported up to 
$z \!\sim\! 0.8$ by using the data of various galaxy surveys 
(\citealt{Tadros1999, Percival2004, Cole2005, Guzzo2008, Blake2011, Samushia2012, 
Reid2012, Beutler2013, delaTorre2013, Contreras2013a, Oka2013}). 
In the near 
future we expect more RSD measurements at even higher redshifts. 
Although the statistical significance is not as large as those at 
lower redshifts, an RSD measurement at $z \!\sim\!3$ has also been 
reported by \citet{Bielby2013}. Planned/on-going surveys, such as 
VLT/VIPERS \footnote{\url{http://vipers.inaf.it/}} ($z \lesssim 1$), 
Subaru/FastSound 
\footnote{\url{http://www.kusastro.kyoto-u.ac.jp/Fastsound/}} ($z \!\sim\! 1.3$) 
and HETDEX \footnote{\url{http://hetdex.org/}} ($z \sim 3$), will give 
further constraints on the modified gravity theories proposed to 
explain the accelerated cosmic expansion.  

However, there are several effects that could result in systematic 
errors of the growth rate measurement, e.g., the non-linear evolution 
of the power spectrum, and the galaxy/halo bias. These must carefully 
be examined in advance of future ambitious surveys, in which the 
systematic error might be larger than the statistical error. 

\cite{Okumura2011} demonstrated the importance of non-linear 
corrections to the growth-rate parameter measurement 
by using the multipole moment method for the linear power spectrum
\citep*{Cole1994} with an assumption of a scale-independent constant
halo bias, by using halo catalogues from N-body simulations at $z \sim
0.3$.  A simple step to go beyond the linear-theory formula is to
include the effect of the velocity dispersion that erases the apparent
fluctuations on small scales \citep{Fisher1994, Peacock1994,
  Hatton1998, Peacock1999, Tinker2006}.  Although this effect was
originally discussed to describe the random motions of galaxies inside
a halo and usually referred to as the Finger-of-God (FoG) effect 
(\citealt{Jackson1972, Tully1978}), 
the presence of any pairwise velocity between galaxies (or even haloes)
results in the damping of the clustering amplitude \citep[see,
e.g.,][]{Scoccimarro2004}.  This is often phenomenologically modeled
by multiplying a damping factor that reflects the pairwise velocity
distribution function.  \cite{Bianchi2012} found that the RSD
parameter $\beta$ measured using this approach has a systematic error
of up to 10 per cent for galaxy-sized haloes in simulated halo
catalogues at $z\!=\!1$.

Another step to include the effect of the non-linear evolution is to 
use analytical redshift-space formulae of the power spectrum and/or 
the correlation function for modestly non-linear scales larger than 
the FoG scale (\citealt{Scoccimarro2004}; \citealt*{Taruya2010} (TNS); 
\citealt{Nishimichi2011, Tang2011, Seljak2011, Reid2011, Kwan2012}). 
\cite{delaTorre2012} showed that an accuracy of 4 per cent is 
achievable for measurements of $f$ from two-dimentional (2D) two-point 
correlation functions, when the TNS formula for the matter power 
spectrum is applied. In these previous studies, the halo 
bias was treated as a constant free parameter, or the correct 
scale-dependence of the bias parameter directly measured from 
numerical simulations was used, to derive the RSD parameters. 
However, in real surveys the true bias cannot 
be measured and hence it is uncertain whether this accuracy can 
really be achieved.  A more practical method to include the effect 
of a general scale-dependent bias is to use phenomenological 
and parametrized bias models, such as the parametrization 
proposed by \citet{Cole2005} (we call it `Q-model bias' in this paper), 
but such models have not been extensively tested in the previous studies. 

In addition to these analytical approaches, there are fully empirical 
RSD models based on N-body simulations both in Fourier and in 
configuration spaces.  \citet{Jennings2011b} reported that, by 
employing their fitting formula for the non-linear power spectra of 
velocity divergence \citep{Jennings2011a}, they can recover the 
correct growth rate $f$ from the redshift-space matter power spectrum. 
Also, \citet{Contreras2013b} developed an empirical fitting function 
of the 2D correlation function, and also recover the correct value of 
the growth rate $f$ from halo catalogues by excluding small-scale 
regions from their analysis. 

In this study, we investigate the accuracy of the RSD measurement 
for various halo catalogues at three redshifts of 0.5, 1.35 and 2. 
Especially, we investigate how the accuracy improves by using 
the TNS formula of the power spectrum with the scale-dependent 
Q-model bias.  We run high-resolution cosmological N-body simulations 
of collisionless dark matter particles, and produce 40 realisations of 
halo catalogues in a comoving volume of 
$3.4 \times 10^8$ $h^{-3}\,\mathrm{Mpc^3}$ at each of the three redshifts. 
We then measure the growth rate $f\sigma_8$ by fitting the 2D halo 
power spectrum $P_\mathrm{halo}(k,\mu)$ with theoretical models, 
where $k$ is the wavenumber and $\mu$ the cosine of the angle 
between the line-of-sight and the wavevector. We search six 
model parameters: $f$, the three parameters of the Q-model bias, 
the one-dimensional velocity dispersion $\sigma_\mathrm{v}$, 
and the amplitude of the mass fluctuations $\sigma_8$.  
The other cosmological parameters are fixed in this study.

This paper is organized as follows. In Sec.~\ref{section:mock}, 
we describe the N-body simulations, the generation of halo 
catalogues, and the measurement of the 2D power spectrum  
$P_\mathrm{sim}(k,\mu)$ for matter and haloes. 
In Sec.~\ref{section:rsdmodel}, we introduce the theoretical RSD 
models that we test, and the Markov-chain-Monte-Carlo (MCMC) 
method with which we measure the systematic and statistical errors 
on $f\sigma_8$ and the other model parameters. We give the 
main results in Sec.~\ref{section:results} with some implications 
for future surveys, and Sec.~\ref{section:conclusions} is devoted to 
the summary of this paper. 

Throughout the paper, 
we assume a flat $\Lambda$CDM cosmology 
with the matter density $\Omega_\mathrm{m}\!=\!0.272$, 
the baryon density $\Omega_\mathrm{b}\!=\!0.046$, 
the cosmological constant $\Omega_\Lambda\!=\!0.728$, 
the spectral index of the primordial fluctuation spectrum 
$n_\mathrm{s}\!=\!0.97$, $\sigma_8\!=\!0.81$, and the Hubble parameter 
$h\!=\!0.70$, which are consistent with the 7-year WMAP results 
\citep{Komatsu2011}.

\section{Mock Catalogue Generation and Power Spectrum Measurement}
\label{section:mock}

In this section, we describe the details of our N-body simulation and 
how to measure the 2D power spectra for matter and haloes.  Although 
our main interest is on the analysis for halo catalogues, we also 
analyse the matter power spectra to check the consistency between 
theoretical predictions and the measured power spectra from 
simulations, and to check if we can measure $f\sigma_8$ correctly when 
the halo bias does not exist. 

\begin{figure}
 \begin{center}
  \includegraphics[width=79mm, angle=0]{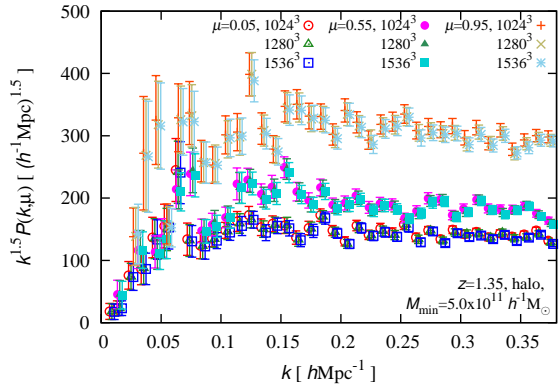}
 \end{center}
 \caption{ 
   The measured 2D power spectra in redshift space for halo catalogues of 
   $M_\mathrm{min}\!=\!5.0\!\times\!10^{11}\,h^{-1}\mathrm{M}_\odot$ 
   at $z\!=\!1.35$. Open, filled and plus (cross, star) symbols show the power 
   spectra at $\mu=0.05,\, 0.55$ and 0.95, respectively. 
   For the same $\mu$ value, three types of points show the power spectra 
   for different mass resolution simulations with  
   $N_p\!=\!1024^3,\, 1280^3$ and $1536^3$ from left to right, respectively.
   Error bars show FKP error estimated as 
   $(\hat{P}+P_\mathrm{shot})/\sqrt{N_\mathrm{mode}}$
   \citep*{Feldman1994}.
   All the data points are on the same $k$ grids but 
   they are slightly shifted horizontally around the true $k$ values for clarity. 
}
 \label{fig:resolution_check}
\end{figure}

We use the cosmological simulation code {\tt GADGET2} 
\citep{Springel2001b, Springel2005}. We employ $N_\mathrm{p}\!=\!1024^3$ dark matter 
particles in cubic boxes of a side length $700\,h^{-1}\mathrm{Mpc}$ 
(or equivalently, a survey volume $V\!\sim\!3.4\!\times\!10^8\,h^{-3}\mathrm{Mpc}^3$) 
with periodic boundary conditions,  
giving the mass resolution of $2.4\!\times\!10^{10}\,h^{-1}\mathrm{M}_\odot$.
This box size is appropriate to achieve the halo mass resolution 
for galaxy surveys. 
The gravitational softening 
length is set to be 4 per cent of the mean inter-particle distance. 
In our simulation, {\tt GADGET2} parameters regarding force and time 
integration accuracy are as follows:
${\tt PMGRID}\!=\!2048^3,\,  {\tt MaxSizeTimestep}\!=\!0.03,\, 
{\tt MaxRMSDisplacementFac}\!=\!0.25$ and ${\tt ErrTolForceAcc}\!=\!0.001$. 
We checked if this parameter choice is adequate 
by comparing with more precise simulations 
(i.e., 
${\tt PMGRID}\!=\!1024^3$, ${\tt MaxSizeTimestep}\!=\!0.005$, 
${\tt MaxRMSDisplacementFac}\!=\!0.01$ and ${\tt ErrTolForceAcc}\!=\!0.0002$).
We ran these simulations from the identical initial condition used for fiducial run,  
and the measured power spectra from them converge (within statistical errors).
In addition, we ran higher mass resolution simulations employing 
$N_\mathrm{p}\!=\!1280^3$ and $1536^3$ particles.
We found that the difference of the power spectra is negligible to 
(see Fig.~\ref{fig:resolution_check}). 
We confirmed that systematic error of the growth-rate measurement 
arising from these changes is smaller than the statistical error.

We generate the initial conditions at $z\!=\!49$ 
using a parallel code developed in \citet{Nishimichi2009} and 
\citet{Valageas2011}, which employs the second-order Lagrangian 
perturbation theory.  The matter transfer function is calculated with 
CAMB (Code for Anisotropies in the Microwave Background; 
\citealt{Lewis2000}).  We run a total of 40 independent realisations 
to reduce the statistical error on the matter and halo power spectra. 
For each realisation, snapshot data are dumped at three redshifts 
$z\!=\!0.5$, 1.35 and 2. 

We identify dark matter haloes using the Friends-of-Friends (FoF) algorithm 
with a linking length $b_\mathrm{FoF}\!=\!0.2$. 
We use a set of halo 
catalogues with different minimum masses in the range of 
$5.0 \times 10^{11}$--$2.0 \times 10^{13}\,h^{-1} \mathrm M_\odot$. 
The detailed properties of the catalogues including the minimum mass 
$M_\mathrm{min}$, the mean halo mass $\overline{M}_\mathrm{halo}$ 
(simple average mass of haloes), and the number density of the 
haloes $n_\mathrm{halo}$ are shown in Table~\ref{table:halo_table}.
Note that, particles grouped into a halo by the FoF algorithm
may include gravitationally unbound ones, in particular for light FoF haloes.
In order to evaluate the effect of this contamination, 
we measured $f\sigma_8$ using only central subhaloes
identified by using {\tt SUBFIND} algorithm \citep{Springel2001b, Nishimichi2013}.
It turns out that this alternative analysis gives consistent $f\sigma_8$ values 
within 1 per cent level with those from the original analysis using FoF haloes.

\begin{table*}
 \caption[catalog]{Summary of the halo catalogues.
  The minimum mass $M_\mathrm{min}$ and the mean halo mass 
  $\overline{M}_\mathrm{halo}$ are shown in units of 
  $h^{-1}\mathrm{M}_\odot$, and the halo number density 
  $n_\mathrm{halo}$ is shown in $h^{3}\mathrm{Mpc}^{-3}$. 
  The halo bias shows the value of 
  $b_0\sigma_8/\sigma_{8,\mathrm{input}}$, where $b_0$ and 
  $\sigma_8$ are the best-fitting parameters by fitting with the 
  TNS+Q-model bias. (See Sec.~\ref{section:rsdmodel_1}
  for the definition of $\sigma_{8,\mathrm{input}}$.) 
 }
\label{table:halo_table}
\begin{center}
\begin{tabular}{cccccccccc} 
\hline
\hline
\multicolumn{1}{c}{}
 & \multicolumn{3}{c}{$z\!=\!2$}
 & \multicolumn{3}{c}{$z\!=\!1.35$}
 & \multicolumn{3}{c}{$z\!=\!0.5$}\\
\hline
$M_\mathrm{min}$
 & $\overline{M}_\mathrm{halo}$ & $n_\mathrm{halo}$ & bias
 & $\overline{M}_\mathrm{halo}$ & $n_\mathrm{halo}$ & bias
 & $\overline{M}_\mathrm{halo}$ & $n_\mathrm{halo}$ & bias \\
\hline
$5.0\times\!10^{11}$
 & $1.51\times\!10^{12}$ & $4.52\times\!10^{-3}$ & 2.3
 & $1.92\times\!10^{12}$ & $6.15\times\!10^{-3}$ & 1.7
 & $2.83\times\!10^{12}$ & $7.43\times\!10^{-3}$ & 1.1\\
$1.0\times\!10^{12}$
 & $2.65\times\!10^{12}$ & $1.91\times\!10^{-3}$ & 2.6
 & $3.32\times\!10^{12}$ & $2.96\times\!10^{-3}$ & 1.9
 & $4.90\times\!10^{12}$ & $3.77\times\!10^{-3}$ & 1.2\\
$2.0\times\!10^{12}$
 & $4.61\times\!10^{12}$ & $7.52\times\!10^{-4}$ & 3.1
 & $5.71\times\!10^{12}$ & $1.28\times\!10^{-3}$ & 2.2
 & $8.36\times\!10^{12}$ & $1.90\times\!10^{-3}$ & 1.4\\
$5.0\times\!10^{12}$
 & $9.80\times\!10^{12}$ & $1.80\times\!10^{-4}$ & 3.9
 & $1.19\times\!10^{13}$ & $3.90\times\!10^{-4}$ & 2.7
 & $1.70\times\!10^{13}$ & $7.22\times\!10^{-4}$ & 1.7\\
$1.0\times\!10^{13}$
 & $1.74\times\!10^{13}$ & $5.14\times\!10^{-5}$ & 4.7
 & $2.08\times\!10^{13}$ & $1.42\times\!10^{-4}$ & 3.3
 & $2.90\times\!10^{13}$ & $3.30\times\!10^{-4}$ & 1.9\\
$2.0\times\!10^{13}$
 & $3.13\times\!10^{13}$ & $1.16\times\!10^{-5}$ & 6.1
 & $3.66\times\!10^{13}$ & $4.43\times\!10^{-5}$ & 4.0
 & $4.96\times\!10^{13}$ & $1.40\times\!10^{-4}$ & 2.3\\
\hline \hline
\end{tabular}
\end{center}
\end{table*}

We measure the 2D power spectra $P_\mathrm{sim}(k,\mu)$ 
for the halo catalogues as well as the matter distribution 
by using the standard method based on the Fourier transform. 
To measure the power spectra in redshift space, the positions 
of haloes (or matter) are shifted along the line-of-sight coordinate 
as ${\bf s}\!=\!{\bf x}+v_z/(aH)\hat{u}_z$ under the plane-parallel 
approximation, where ${\bf s}$ is the redshift-space coordinate, 
${\bf x}$ the real-space counterpart whereas $\hat{u}_z$ 
denotes the unit vector along the line-of-sight.  Then the haloes 
are assigned onto regular $1280^3$ grids through the 
clouds-in-cells (CIC) interpolation scheme, to obtain the density 
field on the grids.  We perform FFT with deconvolution of the 
smoothing effect of the CIC 
\citep{Hockney1988, Takahashi2008, Takahashi2009}. 
We set the wavenumber bin size $\Delta k\!=\!0.01\,h\mathrm{Mpc}^{-1}$ 
and the direction cosine bin size $\Delta \mu\!=\!0.1$. The binned 
power spectrum for a given realisation is estimated as 
\begin{eqnarray}
\hat{P}(k,\mu)
 =\frac{1}{N_\mathrm{mode}}
 \sum_{\mbox{\boldmath{\footnotesize{$k$}}}} 
 \left|\delta_{\mbox{\boldmath {\footnotesize $k$}}}\right|^2-P_\mathrm{shot}
\label{eq:hatP}
\end{eqnarray}
where the summation is taken over $N_\mathrm{mode}$ 
Fourier modes in a bin. In the above equation, $P_\mathrm{shot}$ 
denotes the shot noise given by the inverse of the halo number 
density, $n_\mathrm{halo}^{-1}$, and we do not subtract the 
shot noise for the matter power spectrum. 
We show the measured 2D power spectra $\hat{P}(k,\mu)$ for 
haloes with the mass threshold of 
$M_\mathrm{min}\!=\!5.0\!\times\!10^{11}\,h^{-1}\mathrm{M}_\odot$ at $z\!=\!1.35$ 
in Fig.~\ref{fig:resolution_check}, for three direction cosine values of 
$\mu\!=\!0.05,\,0.55$ and 0.95.
We can see that three power spectra measured from different mass 
resolution simulations (i.e. $N_\mathrm{p}\!=\!1024^3, 1280^3$ and $1536^3$), 
which are started from the same input power spectrum, 
are in good agreement with each other.
Finally, we average 
the 40 independent power spectra and obtain 
$P_\mathrm{ave40}(k,\mu)$ for matter and haloes
\!\footnote{
The measured power spectra, 
both real-space $P^\mathrm{real}(k)$ and redshift-space 2D $P(k,\mu)$, 
are publicly released in \\
\href{http://www.kusastro.kyoto-u.ac.jp/~ishikawa/catalogues/}
{\tt http://www.kusastro.kyoto-u.ac.jp/\~{}ishikawa/catalogues/}
}.

\section{RSD Model Fittings}
\label{section:rsdmodel}
\subsection{Theoretical RSD Models}
\label{section:rsdmodel_1}

In this section we introduce four theoretical models tested in this 
study: two analytical models for the 2D power spectrum in redshift 
space, and two types of parametrization for the halo bias.  We also 
explain how to determine the best-fitting parameters in the models 
through the MCMC method. 

In linear theory, the 2D halo power spectrum in redshift space can 
be written as 
\begin{eqnarray}
P(k,\mu)=b^2(1\!+\! \beta \mu^2)^2P_\mathrm{lin}(k) 
\label{eq:Kaiser_model}
\end{eqnarray}
\citep{Kaiser1987} where $b$ is the halo bias 
and $P_\mathrm{lin}(k)$ the linear matter power spectrum in real space. 
We model the FoG effect arising from halo velocity dispersion by 
the Lorentzian-type damping function: 
\begin{eqnarray}
P(k,\mu)=D_\mathrm{FoG}(k \mu f \sigma_\mathrm{v}) \!\times\! b^2(1 \!+\! \beta \mu^2)^2P_\mathrm{lin}(k)
\label{eq:KaiserFoG_model}
\end{eqnarray}
\begin{eqnarray}
D_\mathrm{FoG}(x)=\frac{1}{(1 \!+\! x^2/2)^2}.
\label{eq:FoG}
\end{eqnarray}
\citep{Peacock1994}.  We call this model `the Kaiser model'. 
As another model that takes into account the non-linear 
evolution on mildly non-linear scales, we use the model based on the 
perturbative expansion \citep{Taruya2010} and generalized to biased 
tracers in \citet{Nishimichi2011}: 
\begin{eqnarray}
\lefteqn{P(k,\mu) = D_\mathrm{FoG}(k \mu f \sigma_\mathrm{v})}\nonumber\\
& \qquad \times b^2\Bigl[
  P_{\delta\delta}(k) \!+\! 2 \beta\mu^2P_{\delta\theta}(k) \!+\! \beta^2\mu^4P_{\theta\theta}(k) 
  \Bigr. \nonumber\\
& \qquad\qquad\qquad\qquad\qquad \Bigl.
  + bC_A(k,\mu;\beta) \!+\! b^2C_B(k,\mu;\beta)\Bigr]
\label{eq:TNS_model}
\end{eqnarray}
where $P_{\delta\delta}$, $P_{\theta\theta}$ and $P_{\delta\theta}$
denote the auto power spectra of density contrast and of velocity divergence 
$\theta \!=\! \nabla\!\cdot\!{\mbox{\boldmath $u$}} 
\!=\! -\nabla\!\cdot\!{\mbox{\boldmath $v$}}/(aHf)$, 
and their cross power spectrum, respectively \citep{Scoccimarro2004,
  Percival2009}, and $C_A$ and $C_B$ are the correction terms arising
from the higher-order mode coupling between the density and velocity
fields \citep{Taruya2010, Nishimichi2011}.  This model is referred to
as `the TNS model' hereafter.  It should be noted that this RSD model
is strictly valid only when the halo bias is assumed to be constant.
However, later we will introduce a scale-dependent halo bias to the
TNS model, to incorporate the scale dependence of bias.  Though there
is an inconsistency here, this is probably the best approach available
for the moment to get a good estimate of $f \sigma_8$.

For our MCMC analysis described in the next subsection, we in advance
prepare templates for the power spectrum of eq.~(\ref{eq:TNS_model})
at each of the three redshifts for a fiducial cosmological model.  In
particular, the three power spectra, $P_{\delta\delta}$,
$P_{\delta\theta}$ and $P_{\theta\theta}$, are calculated by using the
closure approximation up to the second-order Born approximation, and
the correction terms, $C_A$ and $C_B$, are evaluated by the one-loop
standard perturbation theory \citep{Taruya2008, Taruya2009, Taruya2010}.  
In computing these templates, we use the fiducial value of the density 
fluctuation amplitude $\sigma_{8, \mathrm{fid}}(z\!=\!0)\!=\!0.81$ 
and the linear-theory growth factor at each redshift.

In the MCMC analysis, we treat $\sigma_8$ as a free 
parameter and re-scale the template spectra as follows.  We replace 
the density and velocity spectra as 
$P_{ab} \rightarrow P_{ab}\!\times\! (\sigma_8(z)/\sigma_{8,\mathrm{input}}(z))^2$ 
and the correction terms as 
$C_A ({\rm or \ }C_B) \rightarrow C_A ({\rm or \ }C_B) \!\times\!
(\sigma_8(z)/\sigma_{8,\mathrm{input}}(z))^4$. 
These replacements are valid at the leading order, and we 
expect that the error induced by this approximated treatment 
would be small.  This procedure significantly saves computing 
time to calculate the spectra for a given value of $\sigma_8$. 

As for the halo bias, we assume a linear bias 
$b\!=\!\delta_\mathrm{halo}/\delta_\mathrm{matter}$, and we 
adopt two models: a constant bias and a parametrized 
`Q-model' bias to allow scale-dependence (or, equivalently, 
non-locality of the relation between the halo and matter density fields) 
\citep{Cole2005, Nishimichi2011}. These are expressed as 
\begin{eqnarray}
b(k)=\begin{cases}
 \displaystyle{b_0} & \text{: constant bias}\\
 \displaystyle{b_0\,\sqrt{\frac{1+Qk^2}{1+Ak}}} & \text{: Q-model bias} ,
\end{cases}
\label{eq:bias}
\end{eqnarray}
where $b_0$, $Q$ and $A$ are model parameters. 

To summarize, we test the following four theoretical models 
for the 2D halo power spectrum in redshift space: 
`Kaiser+constant bias', `Kaiser+Q-model bias', 
`TNS+constant bias' and `TNS+Q-model bias' 
in this study.  All the models include the four parameters, 
$f, b_0, \sigma_\mathrm{v}$ and $\sigma_8$. 
Additionally, the two models with the Q-model bias have two 
more parameters, $Q$ and $A$. When we analyse the matter 
power spectrum, we fix the bias parameters as 
$b_0 \!=\! 1$, and $Q\!=\!A\!=\!0$.

\subsection{Fitting Methods} 
\label{section:rsdmodel_2}

In this study, we employ the maximum likelihood estimation 
using the MCMC method and find the best-fitting model 
parameters as well as their allowed regions. In contrast to the 
analysis using the ratio of the multipole moments \citep[e.g.][]{Cole1994}, 
we try to fit the shape of the 2D power spectrum, 
$P_\mathrm{sim}(k,\mu)$, directly.  In such a case, we should 
take into account the fact that there is only a small numbers of 
Fourier modes in a $(k,\mu)$ bin. If the measured power spectrum 
$P(k,\mu)$ at each $(k,\mu)$ bin follows the Gaussian distribution, 
the likelihood can be written as $L \propto \exp(-\chi^2/2)$, 
where the chi-square $\chi^2$ is calculated in the standard manner
from the measured and expected values of $P(k,\mu)$ and 
its standard deviation. 

In reality, however, $P(k,\mu)$ does not follow the Gaussian 
but the chi-squared distribution even when the density contrast 
itself is perfectly Gaussian.  In order to take into account this 
statistical property in the maximum likelihood estimation, 
we apply the Wilson-Hilferty (WH) transformation
\citep{Wilson1931} that makes a $\chi^2$ distribution into an 
approximate Gaussian. We define a new variable 
\begin{eqnarray}
P^\prime_\mathrm{sim}(k,\mu)
 =(P_\mathrm{sim} \!+\! P_\mathrm{shot})^{1/3} \ ,
\end{eqnarray}
and $P^\prime_\mathrm{sim}$ is expected to approximately 
obey the Gaussian distribution with a mean of 
\begin{eqnarray}
P^\prime_\mathrm{true} \!=\!\left[1 \!-\! \frac{1}{9 N_\mathrm{mode}}\right]
(P_\mathrm{true} \!+\!P_\mathrm{shot})^{1/3}
\label{eq:WH1}
\end{eqnarray}
and a variance of 
\begin{eqnarray}
\sigma_{P^\prime}^2 \!=\!
\frac{1}{9N_\mathrm{mode}} (P_\mathrm{true} 
\!+\! P_\mathrm{shot})^{2/3}
\ .
\label{eq:WH2}
\end{eqnarray} 
It should be noted that the power spectrum amplitude 
directly measured from the simulations, 
$P_\mathrm{sim}\!+\!P_\mathrm{shot}$, does not exactly obey the 
chi-squared distribution, because it includes the shot noise term. 
However, the WH transformation should be effective only at small 
wavenumbers where the number of modes in a $k$-space bin is small, 
and the shot noise term is relatively unimportant also at 
small wavenumbers. Therefore we adopt the above transformation, 
expecting that $P_\mathrm{sim}\!+\!P_\mathrm{shot}$ approximately 
obeys a chi-squared distribution. (For the wavenumbers where 
the shot noise term becomes comparable with the real-space 
halo power spectrum, see Fig.~\ref{fig:halo_all}.) 

Thus after this transformation, we expect that 
\begin{eqnarray}
\chi^2 =
 \sum_{k<k_\mathrm{max}} \sum_{\mu}
 \frac{ \left[ P^\prime_\mathrm{sim}(k,\mu)-P^\prime_\mathrm{model}(k,\mu)\right]^2}
        {\sigma_{P^\prime\!,\mathrm{model}}^2}
\label{eq:chi-squared}
\end{eqnarray}
approximately obeys a chi-squared distribution, with better accuracy
than simply using $P_\mathrm{sim}$, where $k_\mathrm{max}$ is the
upper bound of the range of wavenumbers that we use in fitting,
$P^\prime_\mathrm{model}$ and $\sigma_{P^\prime_\mathrm{model}}$ are
the WH-transformed model power spectrum and its variance given by
eqs. (\ref{eq:WH1}) and (\ref{eq:WH2}) with replacing
$P_\mathrm{true}$ by the model power spectrum $P_\mathrm{model}$.
In our analyses, we vary $k_\mathrm{max}$ from $0.05$ to
$0.50\,h\mathrm{Mpc}^{-1}$ at an interval of
$0.05\,h\mathrm{Mpc}^{-1}$.

To see how much the fit is improved by this WH approximation, 
we will later compare the results with those obtained using 
the standard chi-square statistic calculation without 
the WH transformation, in which we simply use 
$P_\mathrm{sim}$, $P_\mathrm{model}$ and a variance of 
$\sigma^2_{P}\!=\!(P_\mathrm{model}\!+\!P_\mathrm{shot})^2/N_\mathrm{mode}$
\citep{Feldman1994}. 

Then we find the best-fitting values and their allowed regions of 
all the model parameters (four parameters, $f, b_0, \sigma_\mathrm{v}$ 
and $\sigma_8$, for the models with the constant bias, and additional 
two, $Q$ and $A$, for the models with the Q-model) simultaneously, 
by the standard MCMC technique.

\section{Results and Discussion}\label{section:results}
\subsection{Matter Power Spectrum}\label{section:results_matter}

Before presenting our main results using haloes in the next subsection, 
let us discuss the robustness of the $f\sigma_8$ measurement in the 
absence of the halo/galaxy bias. 

\begin{figure}
 \begin{center}
   \includegraphics[width=80mm, angle=0]{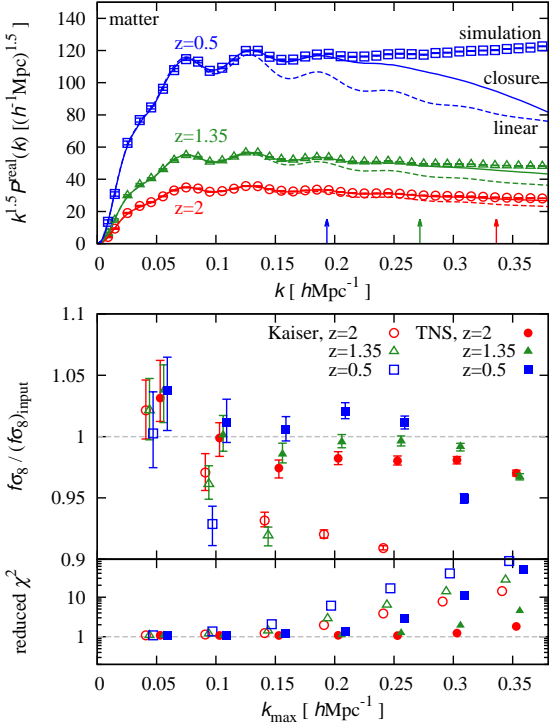}
 \end{center}
 \caption{ Upper panel: Comparison of the measured matter power 
   spectra from simulations, with the theoretical predictions from the 
   closure theory $P_{\delta\delta}$ (solid lines), and from a linear 
   theory $P_\mathrm{lin}$ (dashed lines), at the three different 
   redshifts. Arrows indicate the wavenumbers up to which 
   the closure theory is accurate at 1 per cent level 
   ($k_\mathrm{1\%}\!=\!0.19$, 0.27 and $0.34\,h\mathrm{Mpc}^{-1}$ 
   at $z\!=\!0.5$, 1.35 and 2, respectively). Lower panel: 
   The best-fitting $f\sigma_8$ with 1-$\sigma$ error bars and 
   the reduced chi-squared values by fitting with the Kaiser model 
   (open symbols) and with the TNS model (filled symbols) as a 
   function of the maximum wavenumber $k_\mathrm{max}$ used in 
   fitting.  (All the data points are on the same $k_\mathrm{max}$ grids 
   for the different models and redshifts, but they are slightly shifted 
   horizontally for clarity (see also Fig.~\ref{fig:resolution_check}).}
 \label{fig:matter_check}
\end{figure}

In the upper panel of Fig.~\ref{fig:matter_check}, we show the matter
power spectra in real space at $z\!=\!0.5$, 1.35 and 2 with the
reference wavenumbers $k_{1\%}$, up to which the closure theory is
expected to be accurate within 1 per cent, indicated by arrows
\citep[see][]{Nishimichi2009,Taruya2010}. 
The measured power spectra indeed agree with $P_{\delta\delta}$ 
predicted by the closure theory at $\sim\!3$ per cent level, 
in rough agreement with the definition of $k_{1\%}$.
Therefore we use $k_{1\%}$ as indicators of a few per cent accuracy 
wavenumbers through the paper.
In the lower panel, we show the measured $f\sigma_8$ values normalized
by the correct ones assumed in the simulations, and the reduced
chi-squared values $\chi^2_\mathrm{red}$ for the best-fitting models.
It is clearly seen that $f\sigma_8$ from the Kaiser model (open
symbols) is significantly underestimated at $k_{\max} \gtrsim
0.10\,h\mathrm{Mpc}^{-1}$ at all the redshifts, while the TNS model
(filled symbols) returns $f\sigma_8$ closer to the correct
value, with systematic errors of less than 4 per cent up to $k_{\max}
\sim 0.30\, h \mathrm{Mpc}^{-1}$. As wavenumber increases,
$\chi^2_\mathrm{red}$ boosts up quickly away from unity, and the
maximum wavenumber $k_\mathrm{max}$ up to which $\chi_\mathrm{red}
\simeq 1 $ roughly coincides with $k_{1\%}$.  Systematic overestimates
by the TNS model are seen at $k_\mathrm{max}\!=\!0.20$ and
$0.25\,h\mathrm{Mpc}^{-1}$ at $z\!=\!0.5$, and underestimates at
$k_\mathrm{max}\!>\!0.15 \,h\mathrm{Mpc}^{-1}$ at $z\!=\!2$. The
origin of these is rather uncertain, but these might arise from
sub-percent uncertainty of the power spectrum prediction by
the closure theory, or from the incompleteness in the RSD modeling of
the TNS model.

The MCMC analysis above is done with the power spectrum,
$P_\mathrm{ave40}$, {\it averaged} over $40$ realisations.  Thus, the
number of modes in each of the $(k, \mu)$ bins is rather large
compared with that available in realistic surveys. We therefore
examine the accuracy of the RSD measurement using $\hat{P}$ in
eq.~(\ref{eq:hatP}) for each realisation.

\begin{figure}
 \begin{center}
  \includegraphics[width=80mm, angle=0]{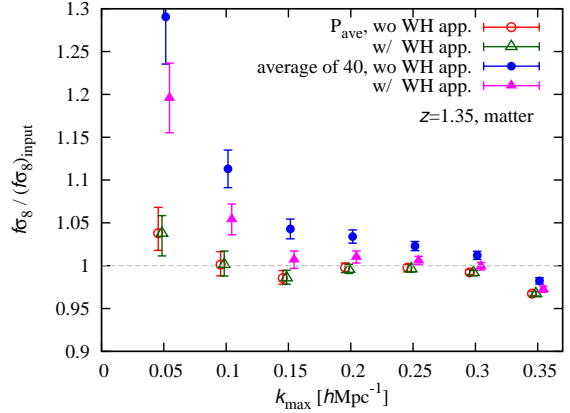}
 \end{center}
 \caption{ Systematic errors of the $f\sigma_8$ measurements by 
   fitting to the matter power spectrum $P_\mathrm{matter}(k,\mu)$ 
   with the TNS model at $z\!=\!1.35$. Open symbols and 
   their error bars show the results from $P_\mathrm{ave40}$ (averaged 
   power spectrum of 40 realisations) and 1-$\sigma$ statistical 
   errors.  Filled symbols show the means of 40 best-fitting $f\sigma_8$ 
   values calculated for each realisation, with the errors estimated by 
   the scatter of the $f\sigma_8$ values of the 40 realisations. Triangles 
   and circles show with and without applying the Wilson-Hilferty 
   approximation, respectively. All the data points are slightly shifted 
   horizontally for clarity
   (see also Fig. \ref{fig:resolution_check}).
   }
 \label{fig:aveof40}
\end{figure}

In Fig. \ref{fig:aveof40} we show by filled symbols the mean values of
the best-fitting $f\sigma_8$ at $z\!=\!1.35$ using the TNS model,
treating each of the 40 realisations as a single observation and
running the MCMC chain for each of them, with and without applying the
WH approximation.  There can be seen overestimations of $f \sigma_8$
at small wavenumbers.  For comparison, we also show the results from
the averaged power spectrum of 40 realisations $P_\mathrm{ave40}$
(open symbols; same as in Fig.~\ref{fig:matter_check}).  Since the
overestimating feature is greatly reduced for the results using
$P_\mathrm{ave40}$ that includes a larger number of modes, the
systematic overestimation must be caused by the small number of modes
in the measured power spectrum. Then we compare the results of
filled symbols with and without the WH transformation (magenta
triangles versus blue circles), and it can be seen that the WH
transformation improves the accuracy of $f \sigma_8$ estimates.  Even
after applying the WH transformation, there still remains a
discrepancy at $k_\mathrm{max} \lesssim 0.10\,h\mathrm{Mpc}^{-1}$, which is
likely to be the limitation of the WH transformation. (Note that the
WH transformation is an approximation.)  However, since the use of the WH
transformation gives more accurate results than those without
using it, this technique is good to be incorporated.

Regarding the sizes of statistical errors on $f\sigma_8$, we also tested jackknife 
resampling method. Although this gives 30--70 per cent larger error bars compared to 
MCMC errors, we think these results are roughly consistent with each other.
In the rest of the present paper, we focus on the results of the MCMC
analyses after averaging over 40 power spectra (i.e.,
$P_\mathrm{ave40}$) with applying the WH transformation, to reduce the
error induced by a small number of modes in $k$-space bins.

\subsection{Halo Power Spectrum}\label{section:results_halo}
\subsubsection{The case of $z=1.35$ and 
$M_{\min}\!=\!1.0 \times10^{12}\,h^{-1}\mathrm M_\odot$}

\begin{figure}
 \begin{center}
   \includegraphics[width=80mm, angle=0]{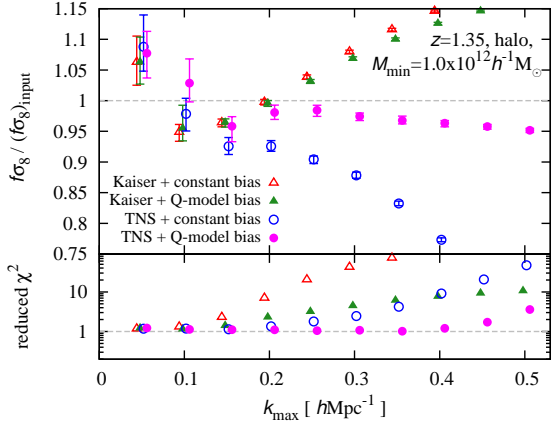}
 \end{center}
 \caption{ The best-fitting $f\sigma_8$ with 1-$\sigma$ error bars 
   and the reduced chi-squared values, for the halo catalogue of 
   $M_{\min}\!=\!1.0\times 10^{12}\,h^{-1} M_\odot$ at $z\!=\!1.35$. 
   Data points show the results of the four different models of the 
   2D halo power spectrum: Kaiser+constant bias, Kaiser+Q-model bias, 
   TNS+constant bias, and TNS+Q-model bias. }
 \label{fig:halo_001_0100_fsigma8}
\end{figure}

We next analyse halo catalogues to measure $f \sigma_8$ 
by fitting the power spectra in redshift space with the four 
analytical models. As the baseline case, we show the measured 
$f\sigma_8$ and the values of $\chi^2_\mathrm{red}$ for the 
best-fitting models to the halo catalogues of 
$M_{\min}\!=\!1.0 \times10^{12}\,h^{-1}\mathrm M_\odot$ at $z\!=\!1.35$ 
in Fig.~\ref{fig:halo_001_0100_fsigma8} 
as a function of the maximum wavenumber, $k_{\max}$, used 
in the analysis. Here and hereafter, when we present results 
for a fixed value of $k_{\max}$, we adopt 
$k_\mathrm{max}\!=\!0.25\,h\mathrm{Mpc}^{-1}$ as the baseline value. 

All the four models give $f\sigma_8$ within a few per cent accuracy at
$k_\mathrm{max}\!\sim\!0.10\,h\mathrm{Mpc}^{-1}$, up to which linear
theory is sufficiently accurate (see dashed lines in the upper panel
of Fig.~\ref{fig:matter_check}).  There can be seen overestimation by
more than 1-$\sigma$ level at
$k_\mathrm{max}\!=\!0.05\,h\mathrm{Mpc}^{-1}$, and they are likely to
be cosmic variances. We have checked that one of the two subsamples
gives $f\sigma_8$ consistent with the correct value within 1-$\sigma$
error when we split the 40 realisations into two groups and analyse
the averaged power spectra of them separately.  On the other hand, 
underestimation at $k_\mathrm{max}\!=\!0.15\,h\mathrm{Mpc}^{-1}$ for
all the models seem to be systematic errors.  It is difficult to
identify the causes of these results, since the measured power
spectrum can be fitted pretty well with reduced $\chi^2$ values of
$\sim\!1$. We leave this issue for future studies.

We then investigate the results from the four RSD modelings one by
one.  The Kaiser model again fails to reproduce the correct
$f\sigma_8$ at $k_\mathrm{max}\!\gtrsim\! 0.25\,h\mathrm{Mpc}^{-1}$, but this
time $f\sigma_8$ are overestimated, in contrast to the results of the
matter power spectra. Even when the TNS model is employed, the
assumption of the constant bias leads to underestimation of
$f\sigma_8$ at $k_\mathrm{max}\!\gtrsim\!0.20\,h\mathrm{Mpc}^{-1}$.
However, when we use the TNS model with the scale-dependent Q-model bias, 
the systematic error is significantly reduced down to 5 per cent level up to 
$k_{\max}\!\sim\!0.50\,h\mathrm{Mpc}^{-1}$.  Note that the adopted
perturbation theory is accurate by $\sim\!1$ per cent level only up to
$k_\mathrm{max}=0.27\,h\mathrm{Mpc}^{-1}$.  It is rather surprising
that the reduced $\chi^2$ values are $\sim 1$ up to
$k_{\max}\!\sim\!0.50\,h\mathrm{Mpc}^{-1}$.  This means that $\sim 5$
per cent level systematic errors of $f \sigma_8$ is possible even if the fit
looks good, which should be kept in mind in future analyses applied on
the real data.

\begin{figure}
 \begin{center}
  \includegraphics[width=80mm, angle=0]{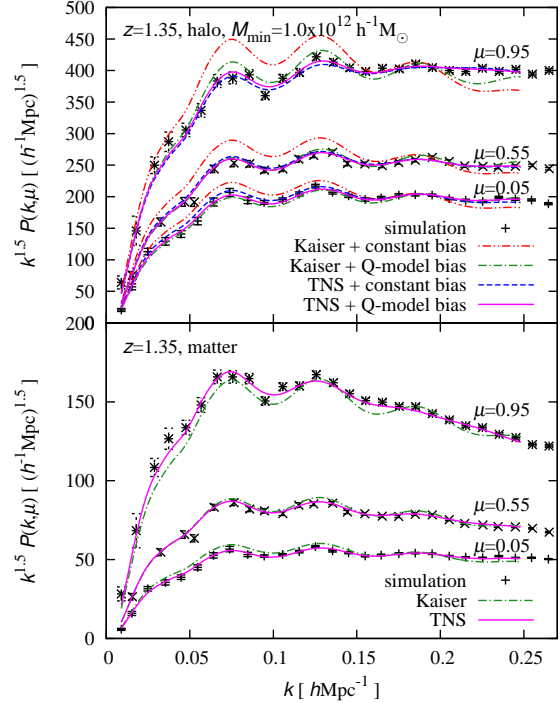}
 \end{center}
 \caption{ The power spectra in redshift space at $\mu\!=\!0.05$, 0.55 
   and 0.95 at $z\!=\!1.35$. The upper panel is for the halo catalogue of 
   $M_{\min}\!=\!1.0 \times 10^{12} h^{-1}\,\mathrm{M}_\odot$, 
   while the lower panel is for the matter distribution. The data points are 
   measurements from simulations and the curves show the best fits for
   different models (see labels in the figure for corresponding models). }
 \label{fig:halo_001_0100_pkmu}
\end{figure}

\begin{figure}
 \begin{center}
   \includegraphics[width=80mm, angle=0]{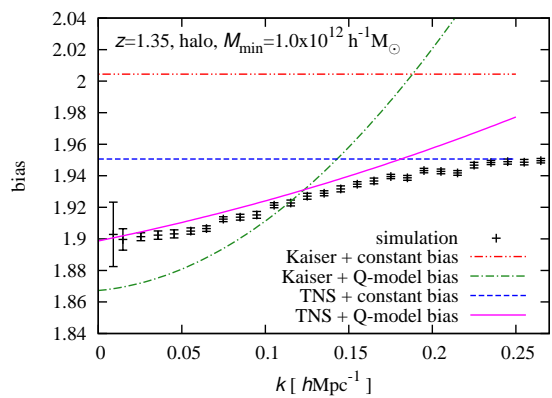}
 \end{center}
 \caption{ Comparison between the halo bias directly measured from 
   simulations and the best-fitting bias models, for the halo catalog of 
   $M_{\min}\!=\!1.0 \times 10^{12} \,h^{-1} \mathrm{M}_\odot$ at $z\!=\!1.35$. 
   The data points and lines are for the same simulation and models 
   as those in the upper panel of Fig.~\ref{fig:halo_001_0100_pkmu}.}
 \label{fig:halo_001_0100_bias}
\end{figure}

We plot in Fig.~\ref{fig:halo_001_0100_pkmu} the four best-fitting 
model power spectra against the simulation data measured at three fixed
direction cosine of the wavevector, $\mu\!=\!0.05$, 0.55 and $0.95$.
In Fig.~\ref{fig:halo_001_0100_bias}, the halo bias measured from
N-body simulations is presented.  The plot shows the mean of the 40
independently-measured biases from each realisation in real space as
$b(k)\!=\!\sqrt{P_\mathrm{halo}(k)/P_\mathrm{matter}(k)}$, and its
standard deviation.  For comparison, we also show the best-fitting
model bias curves, $b(k)\sigma_8/\sigma_{8,\mathrm{input}}$, for the
four models, which are calculated for each model with the
corresponding parameters, $b_0, Q, A$ and $\sigma_8$, using their
best-fitting values found by the MCMC analysis. The measured bias
shows a monotonic increasing trend with the wavenumber. Generally the
scale-dependence of the halo bias is different for different halo mass and
redshift, and both increasing and decreasing trends are possible
depending on these parameters 
(\citealt{Sheth1999, Okumura2011, Nishimichi2011}).

\begin{figure}
 \begin{center}
   \includegraphics[width=80mm, angle=0]{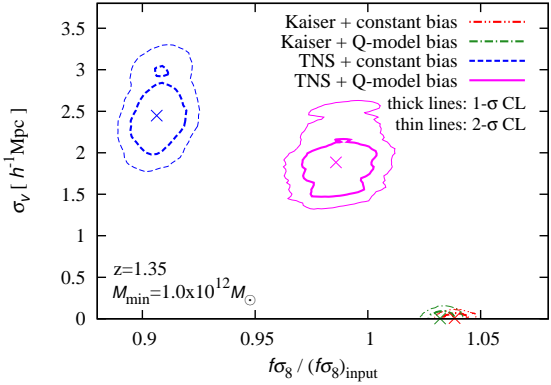}
 \end{center}
 \caption{ The best-fitting values and the 1- and 2-$\sigma$ confidence 
   regions of the four different models (see figure) in the 
   $f\sigma_8$-$\sigma_\mathrm v$ plane, for the halo catalogue of 
   $M_{\min}\!=\!1.0 \times 10^{12} h^{-1} \mathrm{M}_\odot$ at $z\!=\!1.35$.}
 \label{fig:fs8_sv_001_0100}
\end{figure}

When the Kaiser model is used, an apparently inverse trend is seen for 
the systematic deviation of $f\sigma_8$ measurements from the input 
value, for the matter and halo power spectra, and this can be 
understood as follows. In a fitting to the matter spectrum, the Kaiser 
model tries to reproduce the power enhancement arising from the 
non-linear evolution at high-$k$ by setting $\sigma_8$ larger than the 
input value, because of the absence of the bias model parameters 
(see dash-dotted line at $\mu \sim 0$ in the lower panel of 
Fig.~\ref{fig:halo_001_0100_pkmu}). It is easy to show that, from the 
Kaiser formula, a systematically lower value of $f\sigma_8$ than the 
input value is favored to reproduce the RSD effect at large $\mu$, when 
$\sigma_8$ is overestimated. In a fitting to the halo spectrum, there 
are degrees of freedom for bias models, but the non-linear power 
enhancement at high-$k$ cannot be completely absorbed by the constant 
or Q-model bias. The power enhancement can also be absorbed to some 
extent by reducing $\sigma_\mathrm{v}$ in the FoG damping factor, but 
Fig.~\ref{fig:fs8_sv_001_0100} indicates that the best-fitting 
$\sigma_\mathrm{v}$ is zero when the Kaiser model is employed, 
regardless of the bias modelings. The power enhancement that cannot be 
absorbed by bias modelings or the FoG parameter then favors a larger 
$f\sigma_8$ than the correct value, at the cost of a poorer agreement at 
low-$k$. 

The systematic underestimation of $f\sigma_8$ when we employ the
TNS+constant bias model might be a result of the discrepancy between
the correct bias measured directly from simulations and the
best-fitting constant bias at low-$k$ ($k\lesssim
0.15\,h\mathrm{Mpc}^{-1}$, see dashed line in
Fig.~\ref{fig:halo_001_0100_bias}), because the bias shape of the
best-fitting model of the TNS+Q-model bias is close to the
simulation-measured bias.

Compared with the sizes of statistical errors for the Kaiser+constant bias model, 
we get nearly equal sizes of errors for the Kaiser+Q-model bias, 
1.5--2 times larger errors for the TNS+constant bias and 2.5 times larger errors 
for the TNS+Q-model bias. 
The size of statistical error becomes generally larger with increasing
the number of fitting model parameters because of the effect of marginalising, 
though the size of increase is quantitatively different for different models 
because of different ways of parameter degeneracy.

\subsubsection{Dependence on $z$ and $M_{\min}$}
\label{section:z_M_dependence}

\begin{figure}
 \begin{center}
  \includegraphics[width=80mm, angle=0]{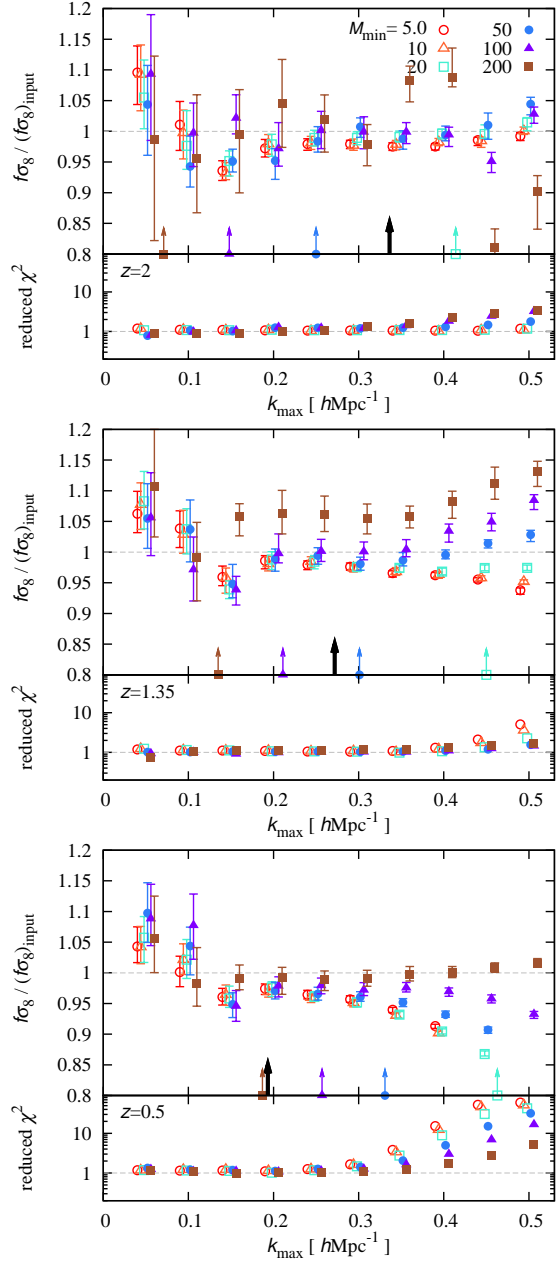}
 \end{center}
 \caption{
  The best-fitting $f\sigma_8$ and the reduced chi-squared values 
  at $z\!=\!2$, 1.35 and 0.5 from top to bottom, respectively. Different 
  symbols are for the different values of the minimum halo mass
  $M_\mathrm{min}\!=\!5.0$, 10, 20, 50, 100 and 200 in units of 
  $10^{11}\,h^{-1}\mathrm{M}_\odot$. Black arrows show the 
  $k_{1\%}$ wavenumbers. Cyan, blue, purple and blown arrows 
  with symbols indicate the wavenumbers where the shot noise term 
  becomes comparable with the halo power spectrum in real space, 
  for the catalogues of $M_\mathrm{min}\!=\!20$, 50, 100 and 200, 
  respectively (see, Sec.~\ref{section:rsdmodel_2}).}
 \label{fig:halo_all}
\end{figure}

Then we investigate the other halo catalogues at the three redshifts
with different minimum halo mass thresholds.  The results of the
$f\sigma_8$ measurement by fitting with the TNS+Q-model bias are shown
in Fig.~\ref{fig:halo_all}.  We firstly focus on the results at
$k_\mathrm{max}\!\sim\!k_{1\%}$. In this regime $f \sigma_8$
measurements with systematic uncertainties of less than $\!\sim\!5$ per cent are
achieved, except for massive halo catalogues of 
$M_{\min}\!\gtrsim\! 10^{13}\,h^{-1}\mathrm{M}_{\odot}$ at $z\!=\!2$. 
These correspond to highly\!
biased haloes of $b_0 \sigma_8/\sigma_{8,\mathrm{input}} \gtrsim\!4$.
Therefore we can state that the TNS model can be used for $f \sigma_8$
measurements with an accuracy of 5 per cent if $k_{\max}\sim k_{1\%}$ 
and $b\lesssim\!4$.

The behavior beyond $k_{1\%}$ depends on the mass of haloes as well as
redshift. In some cases, a value of $f\sigma_8$ consistent with its
input value is successfully recovered up to much higher wavenumbers
(see e.g., the heaviest halo catalogue at $z\!=\!0.5$, from which we
can measure the correct $f\sigma_8$ values up to
$k_\mathrm{max}\!=\!0.45\,h\mathrm{Mpc}^{-1}$). However this result
should be taken with care.  This apparently successful recovery of
$f\sigma_8$ is probably because of the rather flexible functional form
of the scale-dependent bias adopted in this paper. The parameters $A$
and $Q$ can sometimes absorb the mismatch between the true {\it
  matter} power spectra and the TNS model beyond $k_{1\%}$ without
leaving systematics to $f\sigma_8$ for some special cases. The
situation would probably be quite different when different
parameterizations are chosen for $b(k)$.  Nevertheless, it is of
interest to explore the possibility to add some more information from
higher wavenumbers.  Although we, in this paper, employ only one
particular functional form for the scale-dependent bias as well as a
constant bias model, the reproductivity of the growth-rate parameter
from high-$k$ modes with different bias functions is also of
interest. We leave further investigations along this line for future
studies.

\subsection{Implications for Future Surveys}
In this subsection, we give some implications for future use of our
analysis methodologies.  As seen above, we have demonstrated that we
can measure $f\sigma_8$ with a systematic error of $\!\lesssim\!5$ per
cent by using the TNS model combined with the Q-model bias, provided
that the used wavenumber range is $k_{\max}\!\sim\! k_{1\%}$ and haloes
are not strongly biased ($b \lesssim\! 4$).

\cite{Nishimichi2011} showed the expected constraints on the growth 
rate $f(z)$ for some on-going and planned surveys (fig.~6 and table 
III in their paper).  The estimated 1-$\sigma$ statistical errors are 
7.5--3.9 per cent at redshift $z\!=\!0.7$--$1.5$ for 
SuMIRe-PFS\footnote{\url{http://sumire.ipmu.jp/}}, and 5.1 per cent 
at $z\!=\!3.0$ for HETDEX.  This means that the TNS+Q-model bias 
fit can reduce the systematic errors arising from the non-linear 
effects and the halo bias to be comparable or lower than the statistical 
errors from these surveys. 

The space mission Euclid\footnote{\url{http://www.euclid-ec.org/}} will 
survey over a redshift range of $0.7\!<\!z\!<\!2.1$ and get redshifts of 50 
million galaxies.  The number of galaxies in each redshift bin will be 
more than one million.  We can roughly estimate the statistical error 
expected from Euclid to be a few per cent level, by using an 
empirical formula
\begin{eqnarray}
\frac{\Delta f\sigma_8}{\!f\sigma_8}
 \!\sim\! \frac{50}{\sqrt{N_\mathrm{gal}}}
 \label{eq:fs8estimator}
\end{eqnarray}
\citep{Guzzo2008, Song2009} where $N_\mathrm{gal}$ is the 
number of galaxies. This estimation indicates that we need to 
further improve the modeling of RSD, to make the systematic 
error smaller than the statistical error of Euclid.

\section{Conclusions}
\label{section:conclusions}

We have investigated how accurately the structure growth rate 
$f\sigma_8$ can be measured from the RSD effects through 
the model fittings. We have used the halo catalogues generated 
from N-body simulations assuming the standard $\Lambda$CDM 
universe with general relativity, at $z\!=\!0.5$, 1.35 and 2 for 
various minimum halo mass thresholds of 
$5.0 \times10^{11}$-- $2.0 \times 10^{13}\,h^{-1}\mathrm{M}_\odot$.  
We have tested two analytical models for the 2D power spectrum 
in redshift space: the Kaiser model and the TNS model including 
the higher-order coupling terms between the density and velocity fields. 
We have implemented two models for the halo bias: a constant bias 
and a scale-dependent parametric bias model (i.e., Q-model). 

We find significant systematic error (more than 10 per cent for 
$k_{\max}\!\gtrsim\!0.30\,h\mathrm{Mpc}^{-1}$) when the Kaiser model 
is simply adopted regardless of the bias modelings, which is consistent 
with previous studies. Under the assumption of the constant bias, 
the systematic error still remains even when we employ the TNS model. 
However, when we use the TNS model with the Q-model bias, 
the systematic error can be reduced to $\lesssim\!$ 5 per cent for all 
the redshifts and mass thresholds, by using the wavenumber range 
up to $k_{1\%}$ (e.g. $k_{1\%}\!=\!0.19$, 0.27 and 
$0.34 \,h \mathrm{Mpc}^{-1}$ at $z\!=\!0.5$, 1.35 and 2, respectively). 

For some heavy halo catalogues at $z\!=\!0.5$, the TNS+Q-model 
gives the accurate $f\sigma_8$ measurement significantly 
beyond $k_{1\%}$. This is probably because the Q-model bias 
model absorbs the difference between the simulated matter power 
spectrum and the TNS model prediction, but this feature is only 
for particular cases, and a further investigation is necessary. 
At lower mass ranges, the TNS+Q-model gives clearly biased 
$f\sigma_8$ estimates at $k_\mathrm{max}\!>\!k_{1\%}$, especially 
at lower redshifts where the non-linear effects are more significant.

We conclude that the TNS model as a 2D power spectrum formula 
combined with the Q-model bias is a powerful tool to measure 
the structure growth rate. The systematic error can be reduced to 
under 5 per cent at $k_{\max} \!\sim\! k_{1\%}$, which is comparable with 
or smaller than the expected statistical errors of near-future 
ground-based surveys at high redshifts, such as SuMIRe-PFS and 
HETDEX. Some future ambitious surveys, such as Euclid, 
will achieve even smaller statistical errors, and we will need 
to pursue more accurate theoretical models taking into account the 
non-linear effects and the halo/galaxy bias. We also note that 
the TNS formula is valid only when gravity is described 
by general relativity. Therefore this model cannot be used for 
a test of other theories of gravity, but still it can be used to test 
whether general relativity is a valid theory to describe the formation 
of large-scale structure. 

Finally, we note on the importance of an appropriate treatment for 
the $f\sigma_8$ measurement, when only a small number of Fourier 
modes are available in a $k$-space bin of power spectrum measurements. 
In such a case, a measured power spectrum $\hat P$ in 
eq.~(\ref{eq:hatP}) obeys not the Gaussian but the chi-squared 
distribution even when the underlying density field itself obeys the 
Gaussian statistics.  In this study, we have introduced the WH 
transformation which converts the variable obeying the chi-squared 
distribution into an approximate Gaussian, in our likelihood 
calculation of the MCMC analysis.  Indeed we have confirmed that the 
WH transformation improves the accuracy of the $f\sigma_8$ 
measurement, and hence it is a useful prescription
when the number of available modes is small.

\section*{Acknowledgments}
We thank A.~Taruya for providing the templates of the TNS 
power spectrum for our analysis,  and A.~Oka for useful 
discussions. Numerical computations were carried out on 
Cray XT4 and the analyses were in part carried out on 
computers at Center for Computational Astrophysics, CfCA, 
of National Astronomical Observatory of Japan. 
TN is supported by JSPS Postdoctoral Fellowships for Research Abroad. 
RT is supported by Grant-in-Aid for Japan Society for the Promotion 
of Science (No. 25287062) and by Hirosaki University Grant for 
Exploratory Research by Young Scientists. 
NY acknowledge financial support from the Japan 
Society for the Promotion of Science (JSPS) 
Grant-in-Aid for Scientific Research (25287050).

\label{lastpage}

\end{document}